\documentstyle[12pt,epsf,a4]{article}
\begin{document}
\parsep  6pt plus 1pt minus 1pt
\parindent 12truemm
\def\beq{\begin{equation}}
\def\eeq{\end{equation}}
\newcommand{\bea}{\begin{eqnarray}}
\newcommand{\eea}{\end{eqnarray}}
\newcommand{\nn}{\nonumber}
\def\simleq{\; \raise0.3ex\hbox{$<$\kern-0.75em
      \raise-1.1ex\hbox{$\sim$}}\; }
\def\simgeq{\; \raise0.3ex\hbox{$>$\kern-0.75em
      \raise-1.1ex\hbox{$\sim$}}\; }
\def\noi{\noindent}
\font\boldgreek=cmmib10
\textfont9=\boldgreek
\mathchardef\myrho="091A
\def\bfrho{{\fam=9 \myrho}\fam=1}
\def\R{ {\rm R \kern -.31cm I \kern .15cm}}
\def\C{ {\rm C \kern -.15cm \vrule width.5pt
\kern .12cm}}
\def\Z{ {\rm Z \kern -.27cm \angle \kern .02cm}}
\def\N{ {\rm N \kern -.26cm \vrule width.4pt \kern .10cm}}
\def\1{{\rm 1\mskip-4.5mu l} }

\title{Slope of the Isgur-Wise function in the heavy mass limit of quark models \`a la Bakamjian-Thomas}
\maketitle

\centerline{\bf V. Mor\'enas} \par
\centerline{Laboratoire de Physique Corpusculaire} 
\centerline{Universit\'e Blaise Pascal, CNRS/IN2P3, 63177 Aubi\`ere Cedex,
France\footnote{e-mail morenas@clrcls.in2p3.fr}}
\vskip 1 truecm 
 
\centerline{\bf A. Le Yaouanc, L. Oliver, O. P\`ene and J.-C. Raynal} 
\centerline{Laboratoire de Physique Th\'eorique et Hautes Energies\footnote{Laboratoire
associ\'e au Centre National de la Recherche Scientifique - URA D0063\\
\hspace*{\parindent} \phantom{$^1$} e-mail A. Le Yaouanc, leyaouan@qcd.th.u-psud.fr; L. Oliver,
oliver@qcd.th.u-psud.fr; O. P\`ene, pene@qcd.th.u-psud.fr; J-C Raynal,
raynal@qcd.th.u-psud.fr}} \centerline{Universit\'e de Paris XI, b\^atiment 211, 91405 Orsay
Cedex, France} \bigskip

\vskip 1.5 truecm

\begin{abstract}
The slope of the Isgur-Wise function for ground state mesons is evaluated for the heavy mass limit of quark models \`a la Bakamjian-Thomas, which has been previously discussed by us in
general terms. A full calculation in various spectroscopic models with relativistic kinetic energy gives a rather stable result $\rho^2 \approx 1$, much lower than previous estimates. Attention is paid to a careful comparison of this result with the ones of QCD fundamental methods (lattice QCD, QCD sum rules) and with
experimental data.  
\end{abstract}

\vskip 1truecm

\noindent LPTHE Orsay 96-99 \par
\noindent PCCF RI 97-06 \par
\noindent hep-ph/9705324 \par
\noindent December 1996

\newpage
In a series of papers \cite{1} \cite{2} \cite{3}, we have shown that in the heavy mass
limit $m_Q \to \infty$ with one light quark, quark models of form factors based on the old
Bakamjian-Thomas (B-T) construction of relativistic states, and the free quark current, are
presenting very interesting features~: first of all covariance~; but also heavy quark symmetry
for form factors, i.e. heavy to heavy Isgur-Wise scaling  behavior and normalisation
\cite{1} \cite {3}~; duality properties amounting in particular to the Bjorken sum
rule \cite{2}. And finally, we are showing in a forthcoming paper \cite{4} that they
satisfy the new sum rules involving annihilation constants that we have derived in
\cite{5}. \par

Now, of course, such general properties, so interesting as they may be, cannot be the last
word. One has to tackle with concrete, quantitative predictions. This implies making a
specific choice of the mass operator or spectroscopic model, which has been left unspecified
up to now. It is this spectroscopic model which will determine the internal wave functions
at rest of the B-T formalism and lead to the quantitative predictions that we begin to formulate in this letter. \par
We will use actually various possible spectroscopic models that have been presented in the literature ; they are based mainly on the analysis of the spectrum, but indeed the spectrum is still roughly compatible with many possible models. In this situation, the study of transition rates can afford precious additional information and help selecting the spectroscopic models. At this point, one must warn the reader against a possible confusion in the comparison of models. In fact, several authors have proposed at the same time 1) a specific spectroscopic model and 2) a specific formulation of form factors in terms of wave functions, which differ from the B-T formulation ; but the two aspects must be discussed separately. In all the following, it should be clear that we are discussing the results of the various spectroscopic models of these authors within one and the same formulation of form factors, which is the B-T formulation advocated by us, except for some parts where we make also a comparison with the results obtained within the old standard non relativistic formulation of form factors.\par
We begin with
the calculation of a quantity which is the object of both a theoretical and experimental
very active interest, the slope of the Isgur-Wise function at zero recoil, usually
parametrized by~:
\beq
\rho^2 = - {d \xi \over dw}(1) 
\label{1e}
\eeq 
\noi since ${d\xi \over dw}(1) < 0$. The complete calculation of $B \to D,D^*,D^** \ell \nu$ transitions will be presented in a forthcoming paper \cite{6}. The main conclusion of the present paper
is that a proper relativistic treatment of quark models (including the general
Bakamjian-Thomas formulation {\bf and} a relativistic {\bf spectroscopic} model)
gives values of $\rho^2$ around~:
\beq 
\rho^2 \approx 1 \ \ \ .
\label{2e}
\eeq

This value is considerably lower than a previous calculation by Close and Wambach
\cite{close}, who found $\rho^2 = 1.4$. While postponing detailed discussions to the end of
the letter, we want to emphasize from the beginning that our number eq. (\ref{2e}) represents a
very si\-gni\-fi\-cant progress with respect to this previous higher estimate. Indeed, QCD
fundamental methods, i.e. sum rules and numerical lattice QCD, are yielding rather low values.
Sum rules predictions \cite{blok} \cite{neubert1} \cite{bagan} \cite{narison} range from 0.54 to
1. UKQCD lattice calculation \cite{bowler} yields 0.9 as central value, admittedly with very
large error bars \footnote{The first results along the same method were obtained by Bernard, Shen and Soni \cite{bernard}}~:
\beq \rho^2 = 0.9 \matrix{ + 2 \cr
- 3 \cr} {\rm (stat)} \matrix{+ 4 \cr - 2 \cr} {\rm (syst)}. \ \ \
\label{UKQCD}
\eeq

Our quark model prediction $\rho^2 \approx 1$ is therefore much closer to the
predictions of QCD fundamental methods, than the 1.4 of ref. \cite{close}. \par

The progress is even more striking if one considers $s$ quarks instead of $u$ quarks, because
we find a larger difference with \cite{close}, while UKQCD has naturally smaller errors. In
this case, Close and Wambach found $\rho^2_s = 1.64$, while our result is~:

\beq
\rho_s^2 \sim 1.15 \ \ \ . 
\label{3e}
\eeq 
\noi The central value quoted by UKQCD is~:
\beq
\rho_s^2 = 1.2 \matrix{ + 2 \cr
- 2 \cr} {\rm (stat)} \matrix{+ 2 \cr - 1 \cr} {\rm (syst)} \ \ \ .
\label{4e} 
\eeq

It must be kept in mind, as we will explain later, that the results of quark models are not
directly comparable to the renormalization-group invariant $\rho^2$, which is the quantity
given by QCD fundamental methods \footnote{We read the number quoted by Narison in \cite{narison} as "the slope of "physical" Isgur-Wise function" as being actually the renormalisation-group invariant $\rho^2$}, and which we henceforth denote as $\rho^2_{ren.gr.inv.}$. Nevertheless, the progress is quite encouraging. \par

Of course, one would like also to confront theory to experiment, since the predictions of QCD
are not so safe and not so accurate. However, as we will explain, the message of experiment
itself is not so clear. Roughly speaking, the experiment does not determine the slope at zero
recoil as one commonly assumes, but an average slope over the allowed range of $q^2$, which
is sizeably different. Therefore, we prefer to postpone the discussion of
experimental data until the end of the letter. \par

It seems to us, meanwhile, already significant to get agreement with QCD methods, or at least
to get much better compatibility with them than previously obtained in quark models. But
then, how do we obtain this qualitative improvement~? There are two aspects. \par
The main aspect
is that we use relativistic {\bf spectroscopic} models instead of a non-relativistic one
in \cite{close}, i.e. models with a square root kinetic energy $\sqrt{\vec{p}^{\ 2} + m^2}$ instead of $m+\vec{p}^{\ 2}/2m$. One must stress that the necessity of a relativistic expression of form factors (relativistic treatment
of overall hadron motion and relativistic current operator) is quite obvious in calculating $\rho^2$ (otherwise one
would get a very small result $\simleq$~0.3); such a relativistic expression is indeed used by Close and Wambach \cite{close}; but, while adopting such an expression, one can still think of maintaining a non-relativistic
{\bf spectroscopic} model to calculate wave functions at rest, because such models seem to
describe the hadron {\bf spectrum} reasonably \footnote{This distinction between the general treatment of form factors and the spectroscopic models, each of them being possibly either relativistic or non-relativistic, will be present throughout the paper }; this is what has been done by Close and Wambach \cite{close}. What we find however is that {\bf wave
functions} at rest can be strongly different in a relativistic spectroscopic model, although
the spectrum is similar, whence the lowering of $\rho^2$~. \par

Another, more technical, but important aspect is that the very common ap\-pro\-xi\-ma\-tion of
using a variational Gaussian to approximate the wave function \cite{close}, \cite{ISGW},
\cite{jaus}, \cite{grach}, fails even at low recoil, e.g. in the calculation of $\rho^2$,
especially in the context of relativistic spectroscopic equations. Calculating exactly the wave
function results in an appreciably lower estimate of $\rho^2$. \par

To summarize, the improvement we obtain for $\rho^2$ is based on logical ingredients
~: more exact calculation of the wave function, and introduction of
relativistic wave equations, which seems unavoidable since the light quarks in a heavy light
system are strongly relativistic. \par \vskip 5 truemm

\noindent {\bf General statements on $\bf \rho^{\bf 2}$ in the B-T approach}  
\par \vskip 5 truemm

The general expression of $\rho^2$ in our approach is given in ref. \cite{1}~:
\bea
&&\rho^2 = {1 \over 3} \left ( 0\left | p_0 \vec{r}^{\ 2} p_0 \right |0 \right )  + \left ( 0
\left | {2 \over 3} + {1 \over 4} {m^2 \over p_0^2} - {1 \over 3} {m \over p_0 + m} \right |0
\right ) \nn \\
\label{5e}
&& p_0 = \sqrt{\vec{p}^{\ 2} + m^2} \ \ \ . 
\eea 
\noi In the paper ref. \cite{2},
an alternative formula is given, which expresses $\rho^2$ as a sum of three positive contributions, each
having a simple meaning in a small velocity expansion of the matrix element $<v'|\gamma^0|v>$
(normalisation $\delta^3 (p - p')$) 
\bea 
\rho^2 = {1 \over 4} 
+ {1 \over 3} \left ( 0 \left | \left ( {p_0 \vec{r} + \vec{r} p_0 \over 2} \right )^2
\right | 0 \right ) + {1 \over 6} \left ( 0 \left | {\vec{p}^{\ 2} \over (p_0 + m)^2} \right | 0
\right ) \ \ \ .  \label{6e} 
\eea
In both cases the matrix elements denoted by $(0 \left |\ldots \right |0)$ are taken on ground state spatial
internal wave functions ; $\vec{p}$ is the light quark momentum operator, and $\vec{r}$ is the canonically conjugate operator ; i.e., in momentum space which is the most convenient one for this calculation :
\bea
\vec{r}=\ i {\partial \over {\partial \vec{p}}} \ \ \ . \label{defr}
\eea
In eq. (\ref{6e}), the 1/4 corresponds to the contribution of the free
active spin 1/2 heavy quark current to $<v'|\gamma^0|v>$ and it would be the same for any hadron when expressed in terms of the slope of this matrix element, instead of invariant form factors. The two other terms represent effects specific to composite hadrons, and depending on the considered hadron state. The first one   would be present even if quarks would have zero spin. We denote it as $\rho^2_{space}$. The second one comes from the Wigner rotations of the spectator quark spin. We denote it
as $\rho^2_{wigner}$. \par

In \cite{1}, we have found a lower bound on $\rho^2$, larger than  1/4, and specific to
models based on the B-T approach~:
\beq 
\rho^2 \geq {3 \over 4} \ \ \ .
\label{7e} 
\eeq 
\noi As explained in \cite{2}, this lower bound reflects the fact that the expression for
$\rho^2_{space}$ for large $|\vec{p}| \gg m$ is $\rho^2_{space} \sim {1 \over 3} \left ( {|\vec{p}|\vec{r} +
\vec{r}|\vec{p}| \over 2} \right )^2$~; then, from the uncertainty principle $|\vec{p}|r \simgeq 1$, one na\"\i vely expects $\rho^2_{space}$ to be $\simgeq 1$. On the other hand for
$|\vec{p}| \ll m$ (non-relativistic limit), $\rho^2_{space} \sim  {1 \over 3} \ m^2(\vec{r}^{\ 2})$ is very large. \par

For a particular class of internal wave functions, the lower bound on $\rho^2$ must be obviously still larger. For Gaussian wave functions \cite{1}~:
\bea
\rho^2(gaussian) \geq 1.2 \ \ \ . \label{borne}
\eea \par
\vskip 5 truemm

\noindent {\bf Choice of the spectroscopic model} \par \vskip 5 truemm
As our preferred relativistic model to calculate the ground state internal wave function
$\varphi_0(\vec{p}^{\ 2})$, we choose the one of Godfrey and Isgur (GI) \cite{godfrey}.
Certainly, this model is rather complex and because of that, one may not agree on all the
ingredients or assumptions which enter it, or one may find difficult to evaluate their
respective impact on the final result. Nevertheless, what makes the model outstanding is its
covering of the whole spectroscopy, from light to heavy quarks, 

Contrarily to other models in the heavy-light sector, the GI model contains the full set
of spin-dependent forces. For the present purpose of calculating $\rho^2$, we need the wave functions in the
limit $m_Q \to \infty$. In this limit, spin-spin and tensor forces, as well as certain parts of the
spin-orbit forces ( those coming from transverse gluon exchange), disappear.  However, the full set
of spin-dependent forces is required to describe the whole spectrum of states and therefore to assess the validity of the model. The GI model includes relativistic features, among which the root square kinetic energy~:
\beq 
K = \sqrt{\vec{p}_1^{\ 2} + m_1^2} + \sqrt{\vec{p}_2^{\ 2} + m_2^2} \ \ \ ,
\label{8e} 
\eeq
\noi and the above various spin-dependent forces, but also many other effects correcting the gluon exchange potential by momentum dependent factors and smearing. Truely, the latter effects spoil the flavor independence of the potential since they depend on quark masses, and in a somewhat adhoc way. The kinetic energy eq. (\ref{8e}) is shared
by several other models and we will also present the corresponding $\rho^2$ to settle the
discussion.\par

\par \vskip 5 truemm

\noindent {\bf Program of numerical calculation of wave functions} \par \vskip 5 truemm
To calculate the eigenfunctions of the Godfrey-Isgur Hamiltonian, which will be used as the
internal wave functions $\varphi$ in the later calculation of $\rho^2$, we use the
method of these authors, i.e. to diagonalize the matrix of the Hamiltonian, taken on a large harmonic oscillator
basis \cite{godfrey} \footnote{Similar calculations have been done to evaluate the pion and other light meson form factors \cite{cardarelli}}. We have made calculations with basis dimension N as large as 25. As one needs only the $m_Q = \infty$ limit, one can simplify much the program by dropping the spin-dependent contributions vanishing in
this limit. A surprising feature of this limit is that the usual order of $L=1$ levels is reversed, i.e. the $2^+$ is the lowest. We have tested the program by making $m_Q = m_b$ and comparing with the
energy levels given in the paper of Godfrey and Isgur. The agreement is reasonable in view of the above simplifications. In fact, an exact check
of
the program has been obtained in the following way. One has also formulated the
complete program which reproduces exactly the masses of Godfrey and Isgur
at finite quark masses. Then, by dropping the relevant terms, one recovers exactly the numbers obtained with the simplified program. 
Programs and checks have also been made for the other spectroscopic models used in the discussions below. \par \vskip 5 truemm

\noindent {\bf Results for $\bf \rho^{\bf 2}$ and discussion of the wave functions} \par \vskip 5 truemm

Then $\rho^2$ is calculated straightforwardly. We find~:
\beq
\rho_{GI}^2 = 1.023 \ \ \ .
\label{9e}  
\eeq
\noi As announced, this is much lower than the 1.4 result obtained in \cite{close}, using
a relativistic treatment of hadron motion, but a non-relativistic spectroscopic model
for wave functions at rest, namely the ISGW {\bf spectroscopic} model \cite{ISGW}
\footnote{not the ISGW model for {\bf form factors}}. In fact, the result of \cite{close} is obtained with a
Gaussian approximation. Doing a complete numerical treatment for the same model, we obtain
1.283, lower than \cite{close}, but still much higher than our result eq. (\ref{9e}) obtained with the GI spectroscopic model. \par

On the contrary, for spectroscopic equations having a square root kinetic energy like the GI model, but differing by their potential part, as
proposed for example by Veseli and Dunietz(VD) \cite{veseli} which take a linear plus Coulomb potential, or
by Colangelo, Nardulli and Pietroni(CNP) \cite{colangelo}, which use a Richardson potential, we find results
remarkably similar to eq. (\ref{9e})~: resp. $\rho^2 = $~0.98 ; 0.97 \footnote{Let us emphasize that
we borrow from these authors only their {\bf spectroscopic} model and not the
calculation of form factors, which we do in our own way, \`a la Bakamjian-Thomas.}. From the
comparison of all these results, it seems justified to conclude that the decisive reason why
we obtain a low value $\rho^2 \approx 1$, is the use of a {\bf relativistic kinetic
energy} instead of a non-relativistic one in \cite{close}. On the other hand, the result
does not seem very sensitive to the very short distance behavior of the potential, since the
three relativistic spectroscopic models give a similar $\rho^2$, while their Coulomb-like
potential $\alpha_s(r)/r$ has respectively  a regular short distance behavior for GI \footnote{In their paper, $\alpha_s \propto r$, but anyway, the possible potential singularity would be smoothed out by the relativistic "smearing"}, an $1/r$ singular behavior for \cite{veseli} since $\alpha_s = cst$, a milder singularity for \cite{colangelo} since $\alpha_s \sim 1/\ell n r$. Neither is the result sensitive
to the corresponding short distance behavior of $\varphi$, which is respectively regular,
power-like singular, or logarithmically singular when $r \to 0$ \cite{durand}. One must note, however, that, not unexpectedly, the convergence of the numerical method is slower for singular potentials with relativistic kinetic energy, and therefore we have less significant digits in this case. As announced,
the new, low result $\rho^2 \approx 1$ cannot be obtained if one uses the very common
approximation by a Gaussian wave function~:
\beq 
\varphi (r) \propto \exp ( - {\beta^2 r^2 \over 2} ) \ \ \ . 
\label{10e}
\eeq
\noi It is already obvious from the lower bound 1.2 on $\rho^2$ we have found for such wave
functions eq.(\ref{borne}). In fact, choosing $\beta$ variationally, we estimate $\beta \simeq
0.57$ for GI\footnote{Note that this value is quite different from the one quoted for
GI in \cite{jaus} \cite{cardarelli} \cite{grach}.}. From this $\beta$, one would find $\rho^2 \simeq 1.2$
which is sensibly higher than the exact value (+ 17 \%). \par
The stability of the result $\rho^2 \approx 1$ can be understood from the following interesting fact concerning the 
{\bf relativistic wave functions}. The integral giving $\rho^2$
from the momentum space wave function comes essentially from the intervall~: $0.4$ GeV $< p < 2.4$ GeV .
We find that in this interval, for the three {\bf relativistic} models considered, the wave functions are rather close to
each other (see fig. 1) and that they are rather well reproduced by an
exponential form in $r$~:
\bea
 &&\varphi(\vec{r}) = 2a^{-3/2} \exp ( - r/a ) \nn \\
 &&\varphi (\vec{p}) = \sqrt{32 \over \pi} a^{3/2} {1 \over (1 + a^2 \vec{p}^{\ 2})^2} 
\label{11e}
\eea
\noi with $a^{-1} \simeq 0.75$ GeV. Now it is found that for such values of $a$ and $m < 0.3$
GeV, this exponential wave function gives $\rho^2 \approx 1$ (the contribution from outside the
above interval is also small for this wave function). Two other remarks concerning the
behavior in eq. (\ref{11e}) are in order. First the value $a^{-1} \simeq 0.75$ GeV does not
correspond to the expected asymptotic behavior $\exp (- mr)$~; we find that this latter
behavior is only observed at very large distance $r \simgeq 15$ GeV$^{-1}$. Second, our value
is quite in the range observed for wave functions of lattice NRQCD \cite{duncan} when $m_Q \to
\infty$ : correcting the fit of \cite{veseli} for the lattice unit 1.75 GeV  \footnote{This correction is made with agreement of S. Veseli}, we find that the lattice wave function is
quite well reproduced for $0.5 < r < 5$ GeV$^{-1}$ by eq. (\ref{11e}), with $a^{-1} = 0.7$
GeV. On the other hand, it must be emphasized that these relativistic wave functions are quite different from the non-relativistic ones, corresponding to the same mass spectrum. In particular the mean momentum $\sqrt{(0|\vec{p}^{\ 2}|0)} = 0.72$ GeV for GI, is much larger than in the non relativistic ISGW model, where the same quantity is $0.5$ GeV for the $B$ meson ; the large momentum tail is also much higher. The various wave functions considered here are compared graphically in Fig.1.\par
$$\epsfbox{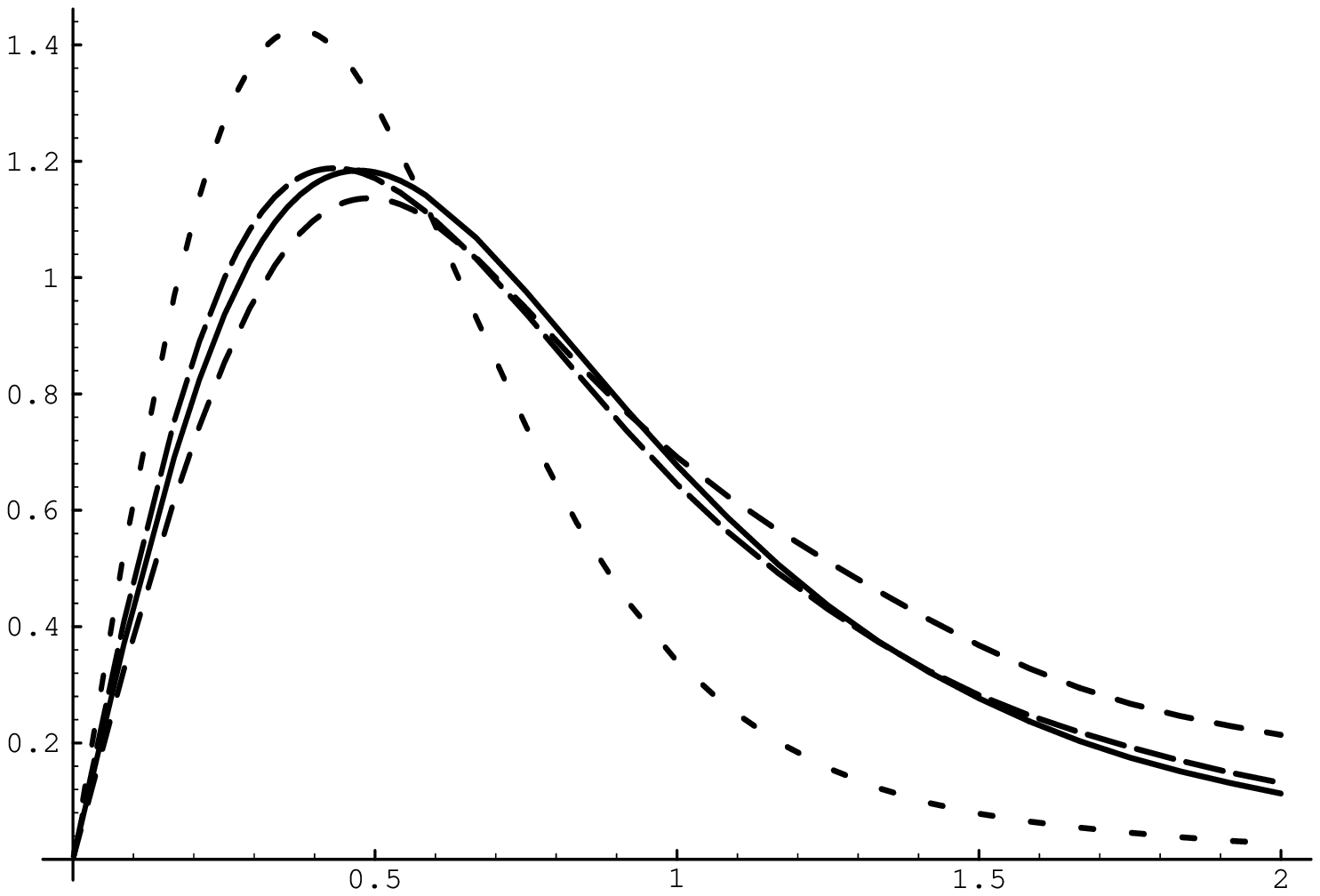}$$
\vskip 5mm

\begin{minipage}[t]{.8 \textwidth}{\small {\bf Fig.1.} Comparison of wave functions in momentum space multiplied by $p$ ;
the integral on $p$ of the squared functions is 1 ; the unit on horizontal axis is 1 GeV ; solid line is for the GI spectroscopic model \cite{godfrey} ; long dashed line
for the CNP model \cite{colangelo} ;
medium dashed line
for the VD model \cite{veseli} ; these three relativistic models are seen to give very similar wave functions ; short dashed line is for the non-relativistic ISGW model \cite{ISGW}, which is seen to yield a much shorter range wave function. } 

\end{minipage}

\vskip 1.5truecm

Finally, let us note that the light quark mass $m$ is not well determined in the above relativistic spectroscopic models. But at the same time $\rho^2$ is not very sensitive to $m$ in
the allowed range (e.g. $m \leq 0.22$ GeV for GI). This is due to the fact that $m$ quite often
enters through the combination $\sqrt{\vec{p}^{\ 2} + m^2}$ and that $m \ll |\vec{p}|$, which corresponds to the fact that the light quark is {\bf ultrarelativistic}. \par \vskip
5 truemm

\noindent {\bf Discussion of the effects of Relativity on $\rho^2$} \par \vskip 5 truemm
 With the wave functions at hand, we can discuss more quantitatively the various relativistic
effects affecting $\rho^2$, which have been discussed in a qualitative manner in \cite{3}
using a formal $v/c$ expansion in the light-quark velocity. We can first compare with the
non-relativistic expression~:
\beq
 \rho^2 = {1 \over 3} m^2 (0| \vec{r}^{\ 2}|0) \ \ \ . 
\label{12e}
\eeq
\noindent With such an expression, a non-relativistic {\bf spectroscopic} model like ISGW
would have given a very low value $\rho^2 = .33 $ ~\footnote{This is the true value, given by the spectroscopic model itself. In the paper \cite{ISGW}, radii are multiplied by a factor 2 to obtain a better fit to empirical radii.},
completely unacceptable. A relativistic {\bf spectroscopic} model of wave functions
combined with the non-relativistic expression eq. (\ref{12e}), if this makes sense, would have
given a still lower and worse value, $\rho^2 = 0.08$ for GI, since the relativistic $(0|\vec{r}^{\ 2}|0)$ is smaller (e.g.
for GI $(0|\vec{r}^{\ 2}|0) = 4.9$ $GeV^2$ versus ISGW $(0|\vec{r}^{\ 2}|0) = 9.1$ $GeV^2$). \par

Now, if we
use, instead of eq. (\ref{12e}), the expression of $\rho^2$ obtained from the relativistic
treatment of {\bf the center-of-mass motion} of B-T, and the relativistic free-quark current, eqs. (\ref{5e}, \ref{6e}), the results are completely changed.
These new relativistic effects (see the general discussion in ref. (\cite{3}) shift $\rho^2$ upwards by a very large amount (around $+ 1$).
This is particularly due to the fact that $\rho_{space}^2$ (second term in the r.h.s. of eq.(\ref{6e})) is increased by the momentum dependent factors $p_0$ with respect to the non-relativistic expression (\ref{12e}). The shift $(\sim + 1)$ of the relativistic expression of $\rho^2$ in terms of given rest
frame wave functions, eqs. (\ref{5e}, \ref{6e}), with respect to the non-relativistic expression, eq.
(\ref{12e}), is in fact much greater than the value of the latter itself. This manifests the
fact that the $v/c$ internal light-quark velocity is not the least small $(\sqrt{(0|\vec{p}^{\ 2}|0)} = .72$ GeV for GI  vs. $m = 0.22$ GeV)
. 
 For
the non-relativistic {\bf spectroscopic} model of ISGW, the result is now so large that it
is again unacceptable. For a relativistic spectroscopic model like GI, the result is
appreciably lower, so that it becomes reasonably compatible with QCD fundamental methods.
\par \vskip 10 truemm

\noindent {\bf Radiative corrections and comparison with QCD fundamental methods} \par \vskip 5
truemm
Although it could  seem more natural to confront
our results first with experimental data, we think that it is in fact a much more involved task than usually believed. In fact, comparison with other theories, although also requiring
a careful discussion, is simpler and more
conclusive, therefore we begin with it. One must warn the reader that many definitions of $\rho^2$ will be used in the following two sections, and that indeed some confusion has arisen about this question in the discussion of quark models. 
When comparing predictions of quark models with the ones of fundamental QCD methods like
numerical lattice QCD or QCD sum rules, one always encounters a basic problem~: the large QCD
radiative corrections are obviously not included in the quark models. In particular quark
models cannot describe the $m_Q \to \infty$ limit of {\bf physical} form factors, since
these should have a logarithmic behavior (log $m_Q)^{\gamma}$ due to the radiative
corrections. These radiative corrections also manifest themselves in the HQET (heavy quark effective theory)
perturbative evolution of the renormalized slope as function
of the renormalisation scale $\mu$. Using the relation given in\cite{neubert2}, and neglecting small $\alpha_s$ terms \footnote{Here and in the following, only leading logarithmic corrections will be given for sake of simplicity, since we aim only at illustrating the effect of radiative corrections}, one has
the general relation~:

\beq 
\rho^2(\mu') = \rho^2(\mu) - {16 \over 81} \ \ln  {\alpha_s(\mu') \over \alpha_s(\mu)}
  \ \ \ .
\label{rhodemu} 
\eeq
\noindent which allows to define also a renormalisation-group invariant slope~:
 
\beq 
\rho^2_{ren.gr.inv.} = \rho^2(\mu) + {16 \over 81} \ \ln  \alpha_s(\mu) \ \ \ .
\label{rhoinv} 
\eeq

 Then, the standard assumption is that quark models describe the $\mu$ dependent
matrix elements, obtained by factorization of the logarithmic coefficients $c_i(m_Q/\mu)$, at
some particular low value of $\mu$, $\mu_0$. In this procedure, it is implicit that, at any
other scale, we shall evolve $\rho^2(\mu)$ according to QCD (more exactly according to HQET in the present case), which amounts to {\bf adding}
the QCD radiative corrections to the brute quark model result. $\mu_0$ could not be much larger
than 1 GeV, since only light-quark degrees of freedom and soft momenta are involved in the $m_Q
\to \infty$ limit of the quark models for $\xi$. On the other hand, if we want perturbation
theory to make sense, we cannot go much lower than a $\mu$ such that $\alpha_s(\mu ) \sim 1$.
This sets bounds for $\mu_0$. With such bounds, we can establish more precisely how to
compare the {\bf renormalization-group invariant} $\rho^2_{ren.gr.inv.}$ of QCD fundamental methods with the
$\rho^2$ of quark models. What we must do is to evaluate in the QCD fundamental methods
$\rho^2(\mu_0)$ which is directly comparable with quark models. \par

We start from the range of values of the renormalization-group invariant slope $\rho^2_{ren.gr.inv.}$, predicted by fundamental
QCD methods, which we denote by an additional QCD subindex. Lattice QCD has very large errors, which by themselves would give a very
large allowed range $0.4 \simleq \rho^2_{QCD,~ ren.gr.inv.} \simleq 1.5$, compatible with almost any estimate. However, taking into account QCD sum rules, we retain the narrower range~:
\beq   0.5 \simleq \rho^2_{QCD,~ ren.gr.inv.} \simleq 1 \ \ \ . 
\label{13e}
\eeq
\noindent Using eq.(\ref{rhoinv})~:
\beq 
\rho^2_{ren.gr.inv.} = \rho^2(\mu_0) + {16 \over 81} \ \ln \alpha_s(\mu_0)
 \ \ \ ,
\label{14e} 
\eeq
\noi with the range of $\mu_0$ indicated above, this leads to~:
\beq
 0.5 \simleq \rho^2_{QCD}(\mu_0) \simleq 1.2 \ \ \ .
\label{15e} 
\eeq

One can conclude that, as announced in the introduction, our quark model prediction $\rho^2 \approx 1$, which is to be compared with $\rho^2_{QCD}(\mu_0)$, is quite compatible with the QCD methods, although closer to the upper edge of the
range. A value like 1.4 \cite{close} seems improbable. \par \vskip 5 truemm

\noindent {\bf Comparison with experiment} \par
\vskip 5 truemm
Comparison with experiment requires a rather detailed discussion. There are two aspects which
complicate the comparison~: the first one is in fact mainly theoretical and has been already
extensively discussed \cite{neubertrep} the second one is purely experimental and has been
perhaps underestimated. \par \vskip 3 truemm

1) The first aspect is that the slopes at $q^2_{max}$ of the physical quantities that are
measured in $B \to D^*\ell \nu$ are not identical with the $\rho$ to be predicted by quark
models. \par

There are three reasons for that~: \par
$\alpha$) the scale corresponding to the physical form factors may be taken around $m_c$,
then, considering the slope of the physical form factor $h_{A_1}$, called $\rho^2_{A_1}$, the
difference with the preceding scale $\mu_0$ , corresponding to the quark model predictions, implies, grossly speaking~:
\beq 
\rho_{A_1}^2 \sim \rho^2(m_c) = \rho^2(\mu_0) - {16 \over 81} \ \ln
 {\alpha_s(m_c) \over \alpha_s(\mu_0)}
\ \ \ . 
\label{16e}
\eeq
\noi This evolution is of course due once more to the {\bf radiative corrections}, and is
based on the assumption that we must {\bf add} these corrections to the quark model brute
prediction, which of course cannot include them. \par
$\beta$) the physical form factor is also affected by $1/m_Q$ corrections, which are not
included in our model, formulated at $m_Q = \infty$. From the discussion of Neubert \cite{neubertrep}, one
concludes that there is a large theoretical uncertainty on these corrections. Nevertheless,
from the estimate of the ``$N_{A_1}$'' ratio, Table 5.1. of Neubert, p. 372 , $N_{A_1} \approx 1$, one may
expect that the $1/m$ corrections largely compensate {\bf numerically} the radiative
corrections in such a way that finally~:
\bea 
\rho^2_{A_1} \sim \rho^2_{ren.gr.inv}  \nn \\
&\left ( = \rho^2(\mu_0) + {16 \over 81} \  \ln  \alpha_s(\mu_0)
\right ) \ \ \ .
\label{17e} 
\eea

$\gamma$) some experiments do not measure directly $\rho^2_{A_1}$ which is the logarithmic
slope of $h_{A_1}$, the main form factor involved in $B \to D^*\ell \nu$, but a quantity
$\widehat{\rho}^2$ related to the slope of $d\Gamma/dw$. According to Neubert,
$\widehat{\rho}^2$ can be safely related to $\rho^2_{A_1}$ according to~:
\beq 
\widehat{\rho}^2 \approx \rho_{A_1}^2 - 0.22 \ \ \ . 
\label{18e}
\eeq
\noi where 0.22 represents $O(1/m_Q)$ corrections due to the contribution of
 form factors other than $h_{A_1}$.Through the above equations, one can a priori 
deduce straightforwardly $\rho^2(\mu_0)$
from the experimental measurements of $\widehat{\rho}^2$ or $\rho^2_{A_1}$, and compare
with the quark model for the range of $\mu_0$ defined above. \par \vskip 3 truemm

2) However, there is still a second aspect, perhaps the main one, which makes the discussion
more complex~: the experiments, in view of the small number of experimental points, do not
actually measure slopes at $w_{min} = 1$, but rather {\bf average slopes} over the
whole range of $w$, $1 < w < 1.5$ \cite{cleo2}. One can make fits which determine the slope
at the origin if one fixes the functional form of the $w$ dependence by a one parameter
function (linear, exponential, ...), but the value of the slope at the origin very strongly
depends on the chosen function~: if the curvature is positive, {\bf the true slope may
be much higher}  than {\bf the usually quoted} {\bf numbers}, which correspond to a linear fit, i.e. to an
average slope ; moreover, the functional form is not strongly constrained by the data ; therefore, the true zero-recoil slope is not well determined.  This phenomenon is generally underestimated when one compares experiment and
theory. In our opinion, the simplest thing to be done is then to compare theoretical and
experimental {\bf average slopes} (average slopes will be denoted by bars). Theoretically, we define such an average slope by
calculating the slope of the straight line
\beq 
\xi_{lin}(w) = 1 - \bar{\rho}^2 (w - 1) 
\label{19e}
\eeq
\noi such that~:
\beq 
\sum_i \left | \xi (w_i) - \xi_{lin}(w_i) \right |^2 
\label{20e}
\eeq    
\noi is minimum for a sufficient large set of points $i$ in the interval $1 < w_i < 1.5$. \par

The result obtained in this way when one uses the GI model for calculating the wave
functions is~:
\beq \bar{\rho}_{GI}^2 = 0.75 \ \ \ . 
\label{21e}
\eeq
\noi It is sizeably lower than the number eq. (\ref{9e}), because the curvature of $\xi$ in our calculation  is positive \footnote{Note that our theoretical results for curvature are not far from the relations proposed
by \cite{caprini}}. In fact, our $\xi$ function is approximately given for the GI case by:
\beq  \xi \approx \left( \frac{2}{1+w} \right)^{2 \rho^2} \ \ \  . 
\label{vincent}
\eeq

As to the experimental numbers, we consider the value of $\rho_{A_1}^2$ \cite{cleo2}, since it is the one which is the most directly related to the
theoretical calculation. The value obtained in the linear fit is in fact to be identified with  $\bar{\rho}_{A_1}^2$, according to eq. (\ref{19e}), and we quote therefore ~:
\beq  \bar{\rho}_{A_1}^2 = 0.91 \pm 0.15 \pm 0.06 \ \ \ . 
\label{experience}
\eeq

As we have explained, the quark model $\rho^2$ is to be identified with
 $\rho^2 (\mu_0)$. Taking into account the relation between $\rho^2 (\mu_0)$ and $\rho^2_{ren.gr.inv.}$, the relation between $\xi_{ren.gr.inv.}$ and
$h_{A_1}$ ($h_{A_1} \sim \xi_{ren.gr.inv.}$) (see eq. \ref{17e}), one should expect, according
to the value of $\mu_0$ :
\beq   \rho_{A_1}^2 \sim \rho^2 (\mu_0) \ {\rm to} \ \rho^2 (\mu_0)+0.2 \ \ \ .
\label{comparaison}
\eeq

\noi Therefore, the comparison between eq. (\ref{21e}) and eq. (\ref{experience}) is quite encouraging. \par

However, the corresponding
result for the ISGW {\bf spectroscopic} model case is $\bar{\rho}_{ISGW}^2 = 0.95$ as
calculated by us ; this is still compatible with
experiment. Therefore, experiment would certainly not exclude presently the latter
{\bf spectroscopic} model\footnote{Let us once more insist that we are not speaking of
the ISGW model of form factors, but of their spectroscopic model, when inserted in the B-T
formalism for form factors}. It seems excluded only by comparison with the QCD methods. A conclusion is that,
due to the phenomenon of the curvature of $\xi$, the present experiments {\bf are not
testing} {\bf accurately} {\bf the slope at the origin}, and they are not
discriminating strongly between mo\-dels through the average slope $\bar{\rho}^2$. \par

Finally, let us note that a similar phenomenon occur at a purely theoretical level, in the analysis of UKQCD lattice results.
The slope $\rho^2$ is estimated through a fit, and the result again depends on the functional
form of the fit. Happily, the dependence is much weaker. The authors have chosen a particular
form to get the result eq. (\ref{4e}). A quadratic fit with unconstrained curvature gives
instead 1 as central value ; according to
us, this latter value is more significant as long
as one has no firm certitude
as regards the magnitude of curvature. \par

In addition, as regards comparison of
fundamental QCD methods with experiment,
it is worth mentioning
that in that case also, one should compare
experiment with the average slope
rather than with the higher slope
at origin.\par \vskip 5 truemm

\noindent {\bf Conclusion} \par
\vskip 5 truemm
In conclusion, we have shown that the relativistic quark models of form factors \`a la
Bakamjian-Thomas, combined with a relativistic spectroscopic model (i.e. with a square root
kinetic energy) to calculate the needed wave functions, yield $\rho^2 \approx 1$. The chosen
spectroscopic model was primarily the Godfrey-Isgur model, but the conclusion was also found
valid for other similar models. This result contrasts with results obtained from a
non-relativistic formulation of form factors and (or) a non-relativistic spectroscopic
models: in the various cases, these results are either much too low or much too high. Our result $\rho^2 \approx 1$ is manifestly in better agreement with QCD fundamental methods, as illustrated by Table 1.
\vskip 5 truemm

\noindent {\bf Acknowledgments} \par
We would like to thank Damir Becirevic and J.-P. Leroy for their help and for many discussions or informations, S.Veseli and I. Dunietz for information on
their calculations, as well as for various explanations, and M. Neubert for explanations on his work.
\newpage 
\begin{center}
\begin{tabular}{|l|c||c|}
\hline
{\bf Quark models} &Form factors : &Form factors : \\
&NR treatment &relativistic treatment (B-T) \\
\hline
NR spectroscopic & & \\
model : ISGW \cite{ISGW} &$\rho^2 =0.33$ &$\rho^2 =1.283$ \\
\hline 
{} &{} &$\rho^2=1.4$~\cite{close}\\
\hline 
{} &{} &$\rho^2_{s} =1.64$~\cite{close} \\
\hline
\hline
Relativistic & & \\ 
spectroscopic models & & \\
\hline
GI~\cite{godfrey} &$\rho^2 =0.08$ &$\rho^2 =$ \quad $1.026^{\bf ~**}$ \\
\hline GI~ &{} &$\rho^2_{s} =$ \quad $1.15^{\bf ~**}$ \\
\hline
VD~\cite{veseli} &$\rho^2 =0.124$ &$\rho^2 =$ \quad $0.98^{\bf ~**}$ \\
\hline
CNP~\cite{colangelo} &$\rho^2 =0.026$ &$\rho^2 =$ \quad $0.97^{\bf ~**}$ \\
\hline
\hline
{\bf QCD fundamental} &\multicolumn{2}{|c|}{}\\
{\bf methods($\rho^2_{ren.gr.inv.}$)} &\multicolumn{2}{|c|}{}\\
\hline
QCD sum rules &\multicolumn{2}{|c|}{$\rho^2 =0.55 \ {\rm to} \ 1$} \\
\hline
Lattice QCD~\cite{bowler} &\multicolumn{2}{|c|}{$\rho^2 = 0.9 \matrix{ + 2 \cr
- 3 \cr} {\rm (stat)} \matrix{+ 4 \cr - 2 \cr} {\rm (syst)}$} \\
{} &\multicolumn{2}{|c|}{$\rho^2_{s} =  1.2 \matrix{ + 2 \cr
- 2 \cr} {\rm (stat)} \matrix{+ 2 \cr - 1 \cr} {\rm (syst)} $} \\
\hline
\end{tabular}
\vskip 5 truemm 
{\bf Table 1.} Theoretical values of $\rho^2$ : quark models, QCD sum rules, lattice QCD.\par
NR stands for non-relativistic. The starred results are the quark model results advocated by us, with both relativistic wave equation and relativistic form factor formulation.
The table illustrates the improvement and stability
gained when including the two features together. All
the results are explained in detail in the text. The
numbers for NR form factor treatment are taken for
conveniency at the B, but should be very close to the true infinite mass limit. The numbers for the relativistic treatment are taken at very large $b$ mass. For quark models, citations refer only to spectroscopic quark models and the $\rho^2$  are the results of our own calculations, except when the
reference is quoted after the value of $\rho^2$. 

\end{center}

\end{document}